\documentclass[letterpaper, preprint, paper,11pt]{AAS}		
\usepackage{bm}
\usepackage{amsmath}
\usepackage[colorlinks=true, pdfstartview=FitV, linkcolor=black, citecolor= black, urlcolor= black]{hyperref}
\usepackage{overcite}
\usepackage{footnpag}			      	
\usepackage{float}

\PaperNumber{21-364}

\usepackage{graphicx}
\usepackage{siunitx}
\title{Autonomous Satellite Detection and Tracking using Optical Flow}

\author{David Zuehlke\thanks{PhD Candidate, Aerospace Engineering, Embry-Riddle Aeronautical University, 1 Aerospace Blvd., Daytona Beach Florida}, 
Daniel Posada\thanks{PhD Student, Aerospace Engineering, Embry-Riddle Aeronautical University, 1 Aerospace Blvd., Daytona Beach Florida}, 
Madhur Tiwari\thanks{PhD Candidate, Aerospace Engineering, Embry-Riddle Aeronautical University, 1 Aerospace Blvd., Daytona Beach Florida}, 
Troy Henderson\thanks{Associate Professor, Aerospace Engineering, Embry-Riddle Aeronautical University, 1 Aerospace Blvd., Daytona Beach Florida}
}

\begin{document}

\maketitle

\begin{abstract}
In this paper, an autonomous method of satellite detection and tracking in images is implemented using optical flow. Optical flow is used to estimate the image velocities of detected objects in a series of space images. Given that most objects in an image will be stars, the overall image velocity from star motion is used to estimate the image frame-to-frame motion. Objects seen to be moving with velocity profiles distinct from the overall image velocity are then classified as potential resident space objects. The detection algorithm is exercised using both simulated star images and ground-based imagery of satellites. Finally, this algorithm will be tested and compared using a commercial and an open-source software approach to provide the reader two different options based on their need.
\end{abstract}

\section{Introduction}
Space situational awareness (SSA) is of increasing importance due to the ever increasing number of objects (both controlled and uncontrolled objects such as debris) in orbit\cite{Virtanen2016-fz,schildknecht_2007}.
In order to guarantee the safety of space assets and personnel in future launches, advancements are needed in the way that current SSA work is performed. Optical measurements offer an inexpensive method of space object tracking and is currently performed by the Space Surveillance Network (SSN). However, the limits of the space surveillance network lie in available time on existing assets for observing satellites.
Utilizing small optical ground stations provides a way to augment the capabilities of the SSN. However, robust tracking and identification methods for data obtained from small stations is a prerequisite to their data being deemed useful\cite{zuehlke_2019}.
Common methods of resident space object (RSO) identification in images involve streak detection, or estimating the gross motion of all objects in the image.\cite{Nir2018-kd,Sease2015-gn} Image template matching has also been utilized as a method of identifying and tracking RSOs in optical imagery\cite{Zuehlke2020}. 

The common thread through the various detection algorithms is that image properties are used to identify and track a specific subset of image features. The computer vision community has further tools to contribute in this arena for a problem framed in the more general terms of tracking the movement of objects in an image (i.e. picking out moving satellites in a star-field).
One of the most widespread methods used for motion tracking in computer vision research today is optical flow\cite{Fortun2015-el}. Essentially optical flow seeks to find the image plane velocity field caused by the motion of the scene or the observer in a sequence of images.
Optical flow has been applied to object tracking in dense object scenes for tracking vehicle motion\cite{xiang2018vehicle} and robot obstacle avoidance and path planning\cite{dur2009optical}. In relation to spacecraft, optical flow has been applied to space-based ground imaging to detect sandstorms, and other large scale weather phenomenon as observed from earth orbiting satellites.\cite{Fortun2015-el} Optical flow has also been used to track the movement of ground objects (such as ships in a harbor) from space\cite{DuChen-2019} and to estimate the angular velocity of a spacecraft using images from a star tracker camera.\cite{Fasano2013-fy} Overall motion of the star-field provided a method of determining the inertial spacecraft angular velocity without any star catalog. However, none of the mentioned research applies optical flow to the problem of space domain awareness, which is the focus of this paper. 

This paper implements an optical flow algorithm based on the Lucas-Kanade method to estimate the camera frame velocities of all objects contained in a space image\cite{Lucas1981-yk}. A space image can be loosely defined as an optical image taken of the night sky that contains stars, astronomical objects, and possibly RSOs. Given that stars make up the bulk of objects in any space image, the overall image velocity provides a way of identifying which objects are stars in the image without the use of a star catalog or plate solving with an application such as Astrometry.net\cite{lang_hogg_2010}.
The velocity estimate for the entire image is taken to be the average velocity of all objects, with the effects of outliers removed. Objects seen to be moving with velocity profiles (magnitude and/or direction) distinct from the overall image motion are classified as potential RSOs. Once all potential RSOs have been identified, the next step in the Space Situational Awareness (SSA) process would be to transform the pixel coordinates to inertial coordinates (angles such as azimuth and elevation or right-ascension (RA) and declination (DEC)) for use in an angles-only orbit determination method. In order to thoroughly test the RSO identification algorithm, a star and RSO image simulator was developed. Simulated imagery provides a testable``ground-truth" scenario. In addition to the simulated imagery, the algorithm is further tested by the use of ground-based images of satellites.

\section{Background and Theory}
\subsection{Optical Flow}
Some preliminary definitions are necessary before delving into the mechanics of optical flow. First, all images can be represented mathematically by a matrix of intensities written as $I(x,y)$. Where $I$ is the intensity at the point $(x,y)$ in the image. The optical flow between two images can be thought of as the 2D vector field representing the apparent motion between two consecutive images \cite{Girosi1989-zv}. As a vector field, optical flow allows the calculation of the displacement and velocities of detected objects in a series of images. There are many methods of computing optical flow, but two of the most common are the methods of Lucas-Kanade and Horn-Schunck \cite{Lucas1981-yk,Horn1981-lk}. In this paper, the Lucas-Kanade (LK) method will be followed. Optical flow assumes:
\begin{enumerate}
    \item The pixel intensities of objects are constant between consecutive frames.
    \item Pixels in a neighborhood all follow similar paths of motion. 
\end{enumerate}
Given these assumptions, a pixel at a time $t$, is represented by $I(x,y,t)$, then the same pixel will be displaced a small amount $(dx,dy)$ when time $dt$ has passed between images. Thus the pixel must obey the relation given in Equation (\ref{eqn:optical:flow}).
\begin{equation}
    I(x,y,t) = I(x + dx,y + dy,t + dt)
    \label{eqn:optical:flow}
\end{equation}
\noindent The optical flow constraint equation is obtained by taking a Taylor series expansion of the right hand side of Equation (\ref{eqn:optical:flow}) \cite{Fortun2015-el}. Performing the expansion we obtain:
\begin{equation}
    I_x u + I_y v + I_t = 0.
    \label{eqn:optical:constraint}
\end{equation}
Where $I_x = \dfrac{\partial{I}}{\partial{x}}$ and $I_y = \dfrac{\partial{I}}{\partial{y}} $ are the spatial image derivatives (image gradients), and $I_t = \dfrac{\partial{I}}{\partial{t}}$ is the temporal image brightness derivative, and $u = \dfrac{dx}{dt}$ is the horizontal optical flow, and finally $v = \dfrac{dy}{dt}$ is the vertical optical flow.
In order to solve Equation (\ref{eqn:optical:constraint}) the LK method takes the original image and divides it into small sections while assuming a constant velocity in the sections. A weighted least-squares equation is then solved to obtain the optical flow fit.
The final solution of the least-squares solution is given by Equation (\ref{eqn:optical:flow:solution}) \cite{Barron1994-vf}. 
\begin{equation}
    \begin{bmatrix}
        u \\
        v
    \end{bmatrix} = \begin{bmatrix}
    \sum_i I_{x_i}^2  & \sum_i I_{x_i} I_{y_i} \\
    \sum_i I_{x_i} I_{y_i} & \sum_i I_{y_i}^2
    \end{bmatrix}^{-1} \begin{bmatrix}
        -\sum_i I_{x_i} I_{t_i} \\
        -\sum_i I_{y_i} I_{t_i}
    \end{bmatrix}
    \label{eqn:optical:flow:solution}
\end{equation}
With the optical flow solution, the velocities of each object detected in consecutive images is obtained.

\subsection{Software Implementation}
To test the effectiveness of this algorithm, two different pipelines that use the LK method will be tested in different development environments. The first environment is set up using MATLAB\textsuperscript{\textregistered}, the second one is performed using the open source toolbox OpenCV for Python.

\subsection{Image Simulator}
\begin{figure}[hbt!]
    \centering
    \includegraphics[width=.45\textwidth]{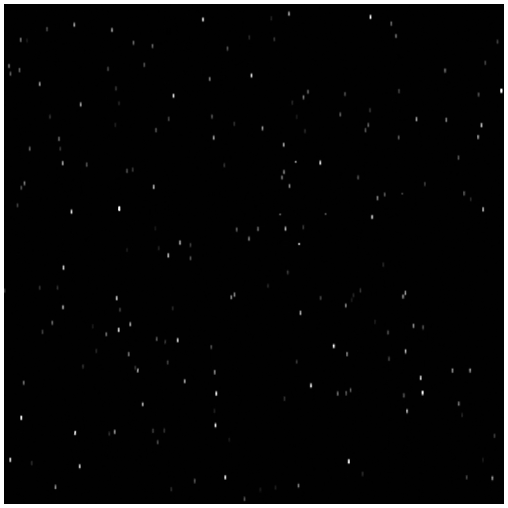}
    \caption{Simulated space image.}
    \label{fig:simulated:space:image}
\end{figure}

In order to test the optical flow tracking method on space images, a star and satellite image simulator was developed. The simulator assumes that the camera frame is tracking at a constant velocity. 
Note that this camera frame velocity can be defined to be the rate of star motion, the rate of RSO motion for a given orbit regime, or simply a constant angular rate.
For the purposes of this simulator, actual star positions were not important, it was desired to simulate the appearance of stars without having to incorporate actual positions from a star catalog.
Star positions are seeded randomly across the image frame, with star brightness levels also set by a random number generator.
To simulate camera motion, each image frame is propagated forward in time for a user defined number of samples.
During each sample the image is convolved with a Gaussian kernel to simulate the effect of an optical system's point spread function (PSF).
Image intensities are scaled according to a user specified camera bit depth. Intensities that fall above the bit depth are automatically set to the bit depth to simulate saturated pixels.
RSO positions are seeded randomly across the image.
RSO velocities are set to be different from the image velocity for stars (as is the case in real images).
Noise is added to each image frame as white Gaussian noise with a mean level set by a user defined percentage of the bit depth.
Further details into the actual implementation of the star image simulator will be provided in the full paper.
Figure \ref{fig:simulated:space:image} shows an example simulated frame with stars that streak slightly in the image and RSOs that show up as point sources. In order to provide more realistic images, Gaussian white noise was added to corrupt the pristine simulated image. White noise was set to be at a value of $5\%$ of the maximum intensity of the image. 

\section{Autonomous RSO Identification Process}
\begin{figure}[hbt!]
    \centering
    \includegraphics[width=.8\textwidth]{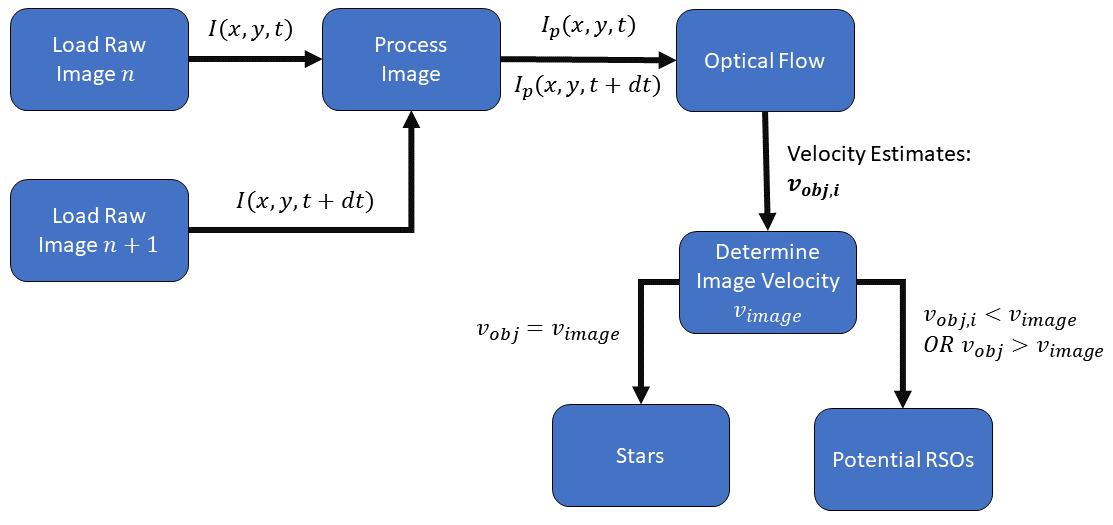}
    \caption{RSO identification process block diagram.}
    \label{fig:optical:flow:process}
\end{figure}

The RSO identification algorithm developed for this research is outlined in Figure \ref{fig:optical:flow:process}. The basic steps shown in Figure \ref{fig:optical:flow:process} are followed by both pipelines.

Given a sequence of consecutive space images, the algorithm begins by loading the image at time $t$ and the image at $ t + dt$, where $t + dt$ is the time of the next image in the sequence. Both images are then processed to reduce noise and improve detection results. The image processing applied involves using a Gaussian smoothing filter to reduce the effects of random image noise, and a threshold filter to remove the background noise level.

Application of a Gaussian smoothing filter suppresses high-frequency content of the image. Gaussian smoothing works by blurring the original image slightly by convolution with a Gaussian Kernel. Each smoothed pixel is a weighted average of the neighboring pixels, with the mean weighted towards the central pixels, thus keeping the image in a Gaussian distribution. A typical Gaussian Kernel such as that given in Equation (\ref{eqn_gaussian}). Where $(x,y)$ represent pixel coordinates and $\sigma$ is the standard deviation for the Gaussian distribution that is desired. A higher $\sigma$ value results in an image that is smoother. 

\begin{equation}
    G(x,y,\sigma) = \dfrac{1}{2 \pi \sigma^2}exp\left(-\dfrac{x^2 + y^2}{2 \sigma^2}\right)
    \label{eqn_gaussian}
\end{equation}
An asterisk represents the convolution operation, then the final smoothed image is given by Equation (\ref{eqn_gauss_conv}) \cite{zuehlke_2019}.
\begin{equation}
    I_s(x,y) = G(x,y,\sigma)*I(x,y) 
    \label{eqn_gauss_conv}
\end{equation}

Applying a threshold on image intensity based on the mean background intensity level provides a simple method of removing any remaining noise after the Gaussian smoothing. All pixels below the chosen threshold level are set to zero intensity, with all remaining illuminated pixels then representing the stars and possible RSOs in an image. The threshold utilized is shown in Equation (\ref{eqn_threshold}). The final processed image, denoted $I_f(x,y)$, is the result of applying the threshold to the image as shown by Equation (\ref{eqn_processed_img}), where only pixel values above the threshold remain in the final processed image. Processed images are then used for the optical flow step. 

\begin{equation}
    \texttt{threshold} = \texttt{mean}(I_s(x,y)) + N \times \texttt{std}(I_s(x,y))
    \label{eqn_threshold}
\end{equation}
\begin{equation}
    I_f(x,y) = I_s(x,y) > \texttt{threshold}
    \label{eqn_processed_img}
\end{equation}

\begin{figure}[hbt!]
    \centering
    \includegraphics[width=.45\textwidth]{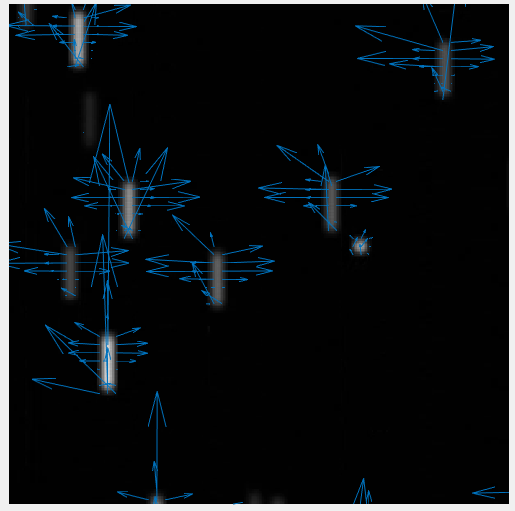}
    \caption{Simulated image optical flow crop example.}
    \label{fig:sim:crop:example}
\end{figure}

Once the processing is applied to both images $I(x,y,t)$ and $I(x,y,t+dt)$, the optical flow between the images is calculated using the Lucas-Kanade method. The result is a velocity field estimating the motion of all objects detected in the images. A cropped image example of the optical flow between two simulated images is shown in Figure \ref{fig:sim:crop:example}.
Now with the velocity field calculated, the next step in the algorithm is to determine the overall image velocity based on the motion of the most prevalent objects in the image (i.e. stars).This is accomplished by looking at the magnitudes of the image velocity for all objects detected in the image.  

\subsection{MATLAB Pipeline}
Once the optical flow between the images has been found, the overall image velocity is calculated based on the magnitude of velocities for image objects. Objects determined to be moving with velocities near the overall image velocity are classified as stars, while objects that move with velocities that are different from the overall image velocity are classified as potential RSOs. 

\begin{figure}[hbt!]
    \centering
    \includegraphics[width=.5\textwidth]{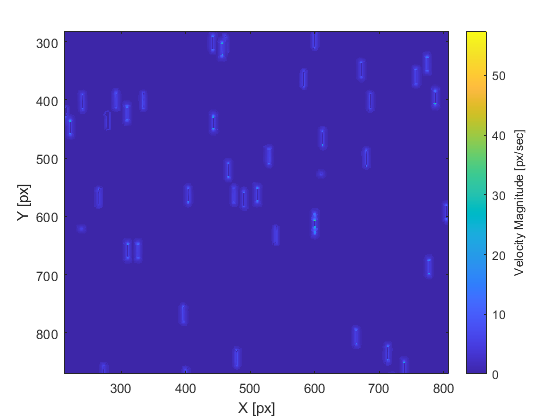}
    \caption{Simulated image optical flow velocity magnitude.}
    \label{fig:sim:magnitude}
\end{figure}

MATLAB includes a direct function to estimate optical flow based on different methods including the LK algorithm. These methods\cite{estimate_object_velocities} include the use of functions and objects such as:
\begin{itemize}
    \item opticalFlowLK: This function creates an object to estimate the direction and speed of a moving feature.
    \item estimateFlow: This function takes the opticalFlowLK object and estimates optical flow between two consecutive video frames. The method runs recursively as the frames in the video advance. 
\end{itemize}
Utilizing the built in functions, the velocity profile of the image is found. Given the velocity magnitude map, we can ``centroid'' the magnitude map as an image whose intensities consist of the individual pixel object velocities. MATLAB's computation of Optical Flow uses a fixed object size and containts one parameter for a noise threshold to determine what is considered a pixel object. By looking at the entire magnitude map then the actual full ``object'' velocity is found rather than just the velocity of a small portion of the object. An example of the resulting velocity magnitude (cropped version) from a simulated image is shown in Figure \ref{fig:sim:magnitude}.
Velocities of each object are computed as the sum of the ``intensity'' for the detected object in the velocity magnitude matrix. With the velocity of all the image objects, now the maximum, minimum, mean, and median velocity objects can be found. It was found that given a sufficient number of stars to set the mean velocity, then by looking at the standard deviation of the velocities RSOs can be picked out as shown in Equations (\ref{eqn:rsovel:plus}) and (\ref{eqn:rsovel:low}). 
\begin{equation}
    v_{rso_k} = v_{mean} + 2 \, \sigma v_i \;\;\;\;\;\;\;\;\;\;i = 1,2,...,N \\
    \label{eqn:rsovel:plus}
\end{equation}
\begin{equation}
    v_{rso_k} = v_{mean} - 2 \, \sigma v_i \;\;\;\;\;\;\;\;\;\; i = 1,2,...,N
    \label{eqn:rsovel:low}
\end{equation}
Where $v_{rso_k}$ is the velocity of the $k^{th}$ rso detected and the index $i$ is the object detected in the current velocity magnitude map with $N$ as the number of total objects detected. 

\subsection{OpenCV Pipeline}
Similar to the MATLAB implementation, the output of the LK algorithm is the position of the different features in terms of pixel location. 
With the time between frames, the velocity of a matched pair between frames can be obtained using a simple derivative of pixels per frame time step:
\begin{equation}
    vel_x = \frac{x_2 - x_1}{frame\,time}\quad \quad vel_y = \frac{y_2 - y_1}{frame\,time}
\end{equation}
Once the velocities are obtained the magnitude of the velocity of the features can be obtained by using the $l_2$ norm.
\begin{equation}
    |vel| = \sqrt{vel_x^2+vel_y^2}
\end{equation}
Once the magnitudes are obtained, a comparison is done between the different magnitudes to filter out the RSOs from the stars. The RSOs have a higher velocity therefore the flow will be higher making them easier to identify. 
It is important to remind the reader that OpenCV is a free toolbox compatible with Python and C/C++ making these approaches completely free for the scientific community.

Similar to MATLAB, OpenCV provides multiple functions\cite{opencv} but the main methods for optical flow estimation are:
\begin{itemize}
    \item goodFeaturesToTrack(): This function creates an object with features to track.
    \item calcOpticalFlowPyrLK(): This function takes the features detected and estimates optical flow between two consecutive video frames.
\end{itemize}
Using these main functions, an algorithm that follows the process outlined in Figure \ref{fig:optical:flow:process} was created. The main difference in the OpenCV implementation lies in the calculation of the object image velocities as mentioned above. Once the image frame velocities are computed then the RSOs can be found from the stars in the image. 

\section{Results}
The autonomous RSO detection method was applied to both simulated and actual imagery. Both the MATLAB and OpenCV pipelines were successfully able to track the RSOs in simulated and real imagery. 
\subsection{Simulated Images MATLAB Results}
\begin{figure}[htb!]
    \centering
    \begin{minipage}[t]{.5\textwidth}
        \centering
        \includegraphics[width=.8\textwidth]{Figures/simulated_star_frame.png}
        \caption{Simulated star field.}
        \label{fig:simulated:space:image:starfield}
    \end{minipage}%
    \begin{minipage}[t]{0.5\textwidth}
        \centering
        \includegraphics[width=.8\textwidth]{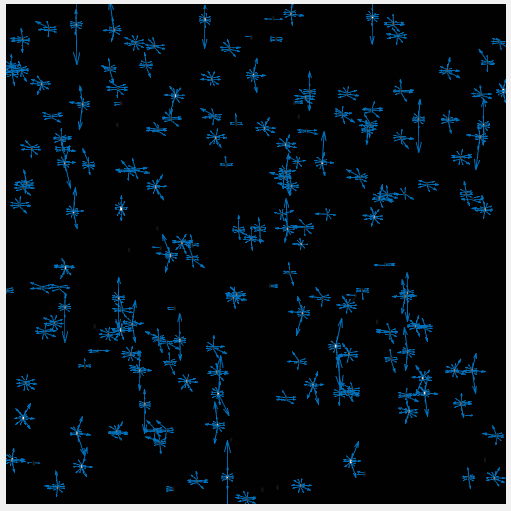}
        \caption{Optical Flow from simulated star field.}
        \label{fig:optical:flow:simulated:example}
    \end{minipage}
\end{figure}

\begin{figure}[h!]
    \centering
    \begin{minipage}[t]{0.5\textwidth}
        \centering
        \includegraphics[width=.8\textwidth]{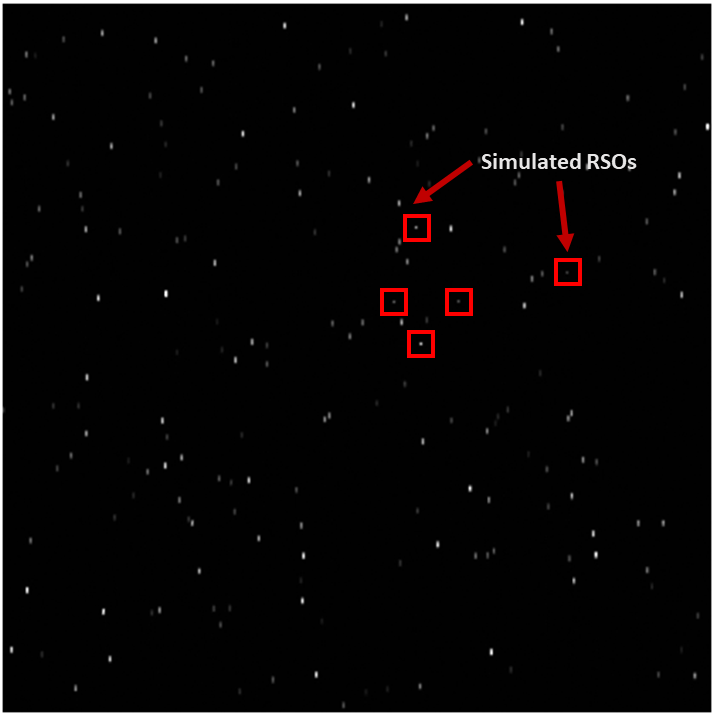}
        \caption{Simulated RSOs highlighted.}
        \label{fig:simulated:RSOs:highlighted}
    \end{minipage}%
    \begin{minipage}[t]{0.5\textwidth}
        \centering
        \includegraphics[width=.8\textwidth]{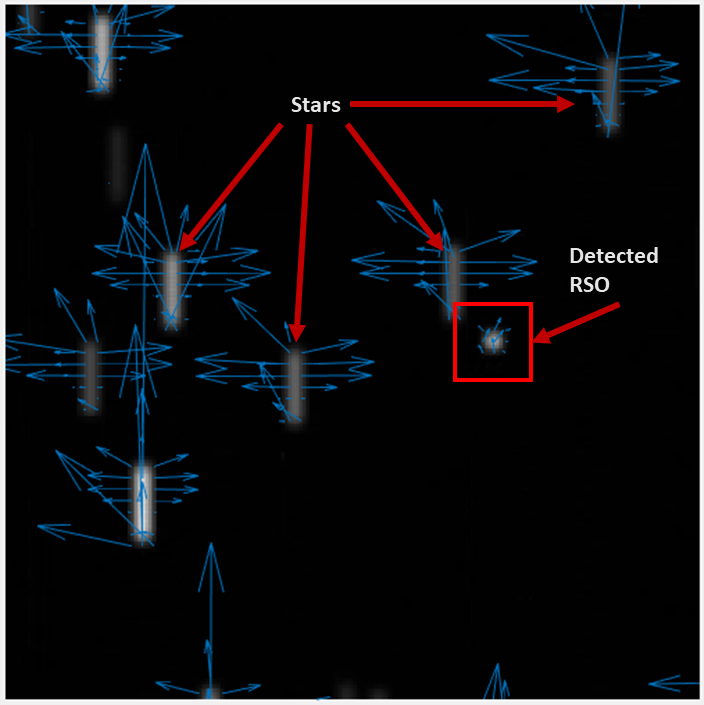}
        \caption{Closeup of simulated image\newline optical flow showing identified RSO.}
        \label{fig:simulated:RSO:identified}
    \end{minipage}
\end{figure}

First, the MATLAB implementation of the The optical flow tracker is shown to be able to successfully track the motion of stars across for both simulated and experimental images. Figure \ref{fig:simulated:space:image:starfield} shows a simulated image star field with a group of 5 stationary RSOs.This simulates a GEO constellation of satellites being tracked. Next, Figure \ref{fig:optical:flow:simulated:example} shows the optical flow for the simulated star field image (computed with the consecutive frame). Blue arrows represent the velocity magnitude and direction estimated for the local pixel regions. Note that the magnitude scale (length of the vectors) is exaggerated for the sake of display. Figure \ref{fig:simulated:RSOs:highlighted} shows the simulated star field with RSOs highlighted. A cropped result of running the optical flow and RSO identification routine is shown in Figure \ref{fig:simulated:RSO:identified}. Note that the stars all have large velocity magnitudes while the detected RSO velocity vectors are much smaller. The algorithm was able to successfully differentiate the stars from the RSOs for the simulated image set. 

\subsection{Experimental Images}
\begin{figure}[htb!]
    \centering
    \begin{minipage}{.5\textwidth}
        \centering
        \includegraphics[width=.85\textwidth]{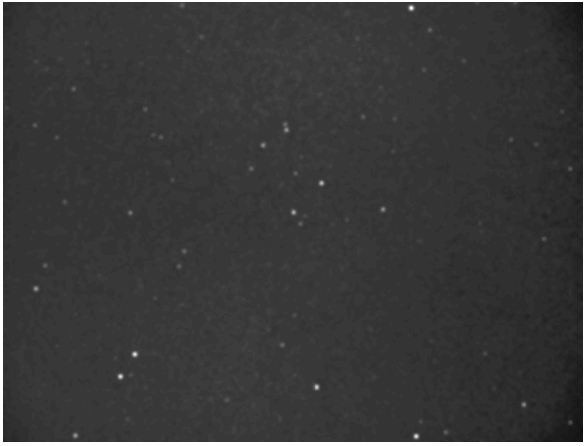}
        \caption{Raw ground-based image.}
        \label{fig:raw:ground:image:matlab}
    \end{minipage}%
    \begin{minipage}{0.5\textwidth}
        \centering
        \includegraphics[width=.85\textwidth]{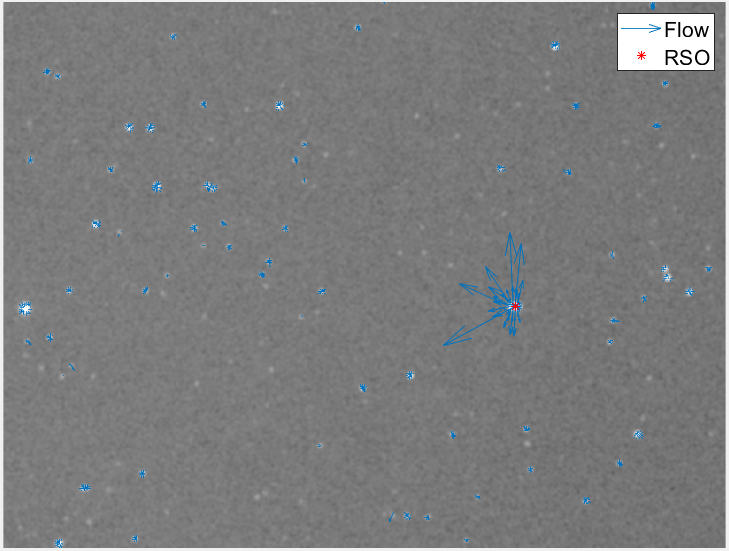}
        \caption{MATLAB optical flow RSO detection.}
        \label{fig:proc:ground:image:matlab}
    \end{minipage}
\end{figure}

\begin{figure}[htb!]
    \centering
    \begin{minipage}[t]{.5\textwidth}
        \centering
        \includegraphics[width=.9\textwidth]{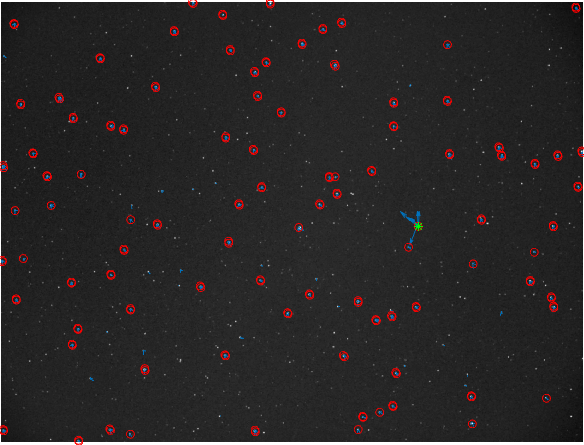}
        \caption{RSO detection frame 10.}
        \label{fig:exp:mat1}
    \end{minipage}%
    \begin{minipage}[t]{0.5\textwidth}
        \centering
        \includegraphics[width=.9\textwidth]{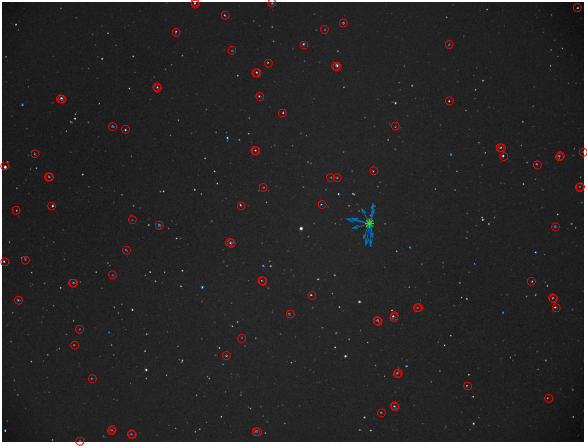}
        \caption{RSO detection frame 30.}
        \label{fig:exp:mat2}
    \end{minipage}
\end{figure}

\begin{figure}[h!]
    \centering
    \begin{minipage}[t]{0.5\textwidth}
        \centering
        \includegraphics[width=.9\textwidth]{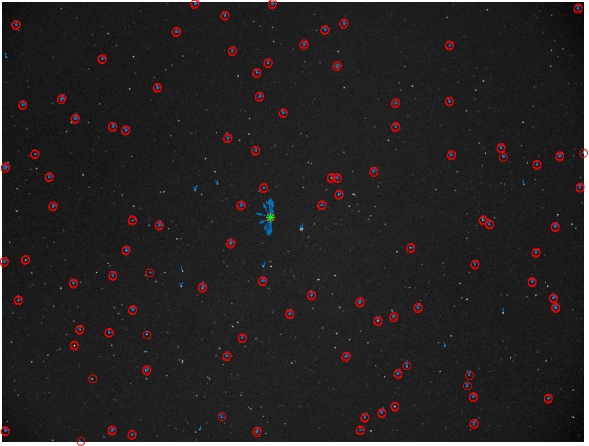}
        \caption{RSO detection frame 50.}
        \label{fig:exp:mat3}
    \end{minipage}%
    \begin{minipage}[t]{0.5\textwidth}
        \centering
        \includegraphics[width=.9\textwidth]{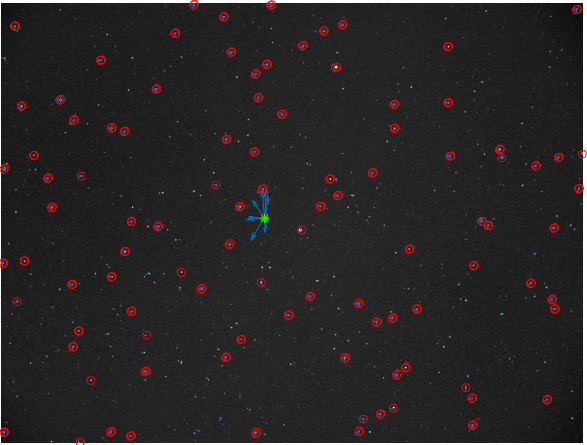}
        \caption{RSO detection frame 70}
        \label{fig:exp:mat4}
    \end{minipage}
\end{figure}

In this section results from running the optical flow algorithm on ground-based imagery obtained from Daytona Beach Florida are presented. Figure \ref{fig:raw:ground:image:matlab} shows a raw space image and Figure \ref{fig:proc:ground:image:matlab} shows the resulting optical flow calculation and RSO detection step. Blue arrows represent the optical flow velocities estimated for various objects detected in the image, and the RSO position is denoted by a red asterisk. This set of images was taken in sidereal tracking mode, so star positions remain nearly constant while the RSO is seen moving through the frame with a much higher image velocity. The algorithm was able to autonomously label the correct object in the image as an RSO. 
Figures (\ref{fig:exp:mat1} - \ref{fig:exp:mat4}) show the results for a longer series of images of the same object. A single RSO was detected moving from the right center of the image towards the center of the image. Note that detected objects are shown in red, the RSO in green, and the flow velocity vectors in blue. 

\subsection{OpenCV Pipeline Results}
\begin{figure}[htb!]
    \centering
    \begin{minipage}[t]{.5\textwidth}
        \centering
        \includegraphics[width=.95\textwidth]{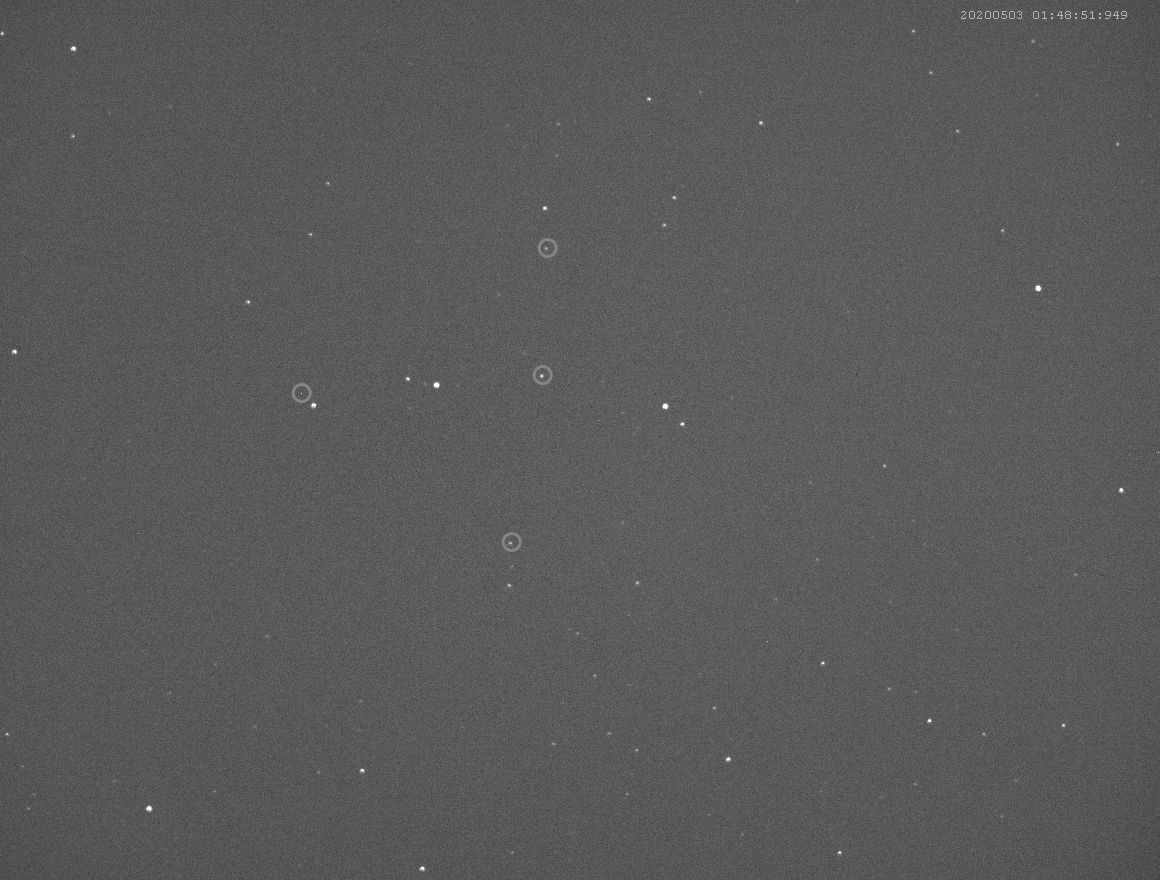}
        \caption{Raw ground-based image.}
        \label{fig:raw:ground:image:opencv}
    \end{minipage}%
    \begin{minipage}[t]{0.5\textwidth}
        \centering
        \includegraphics[width=.95\textwidth]{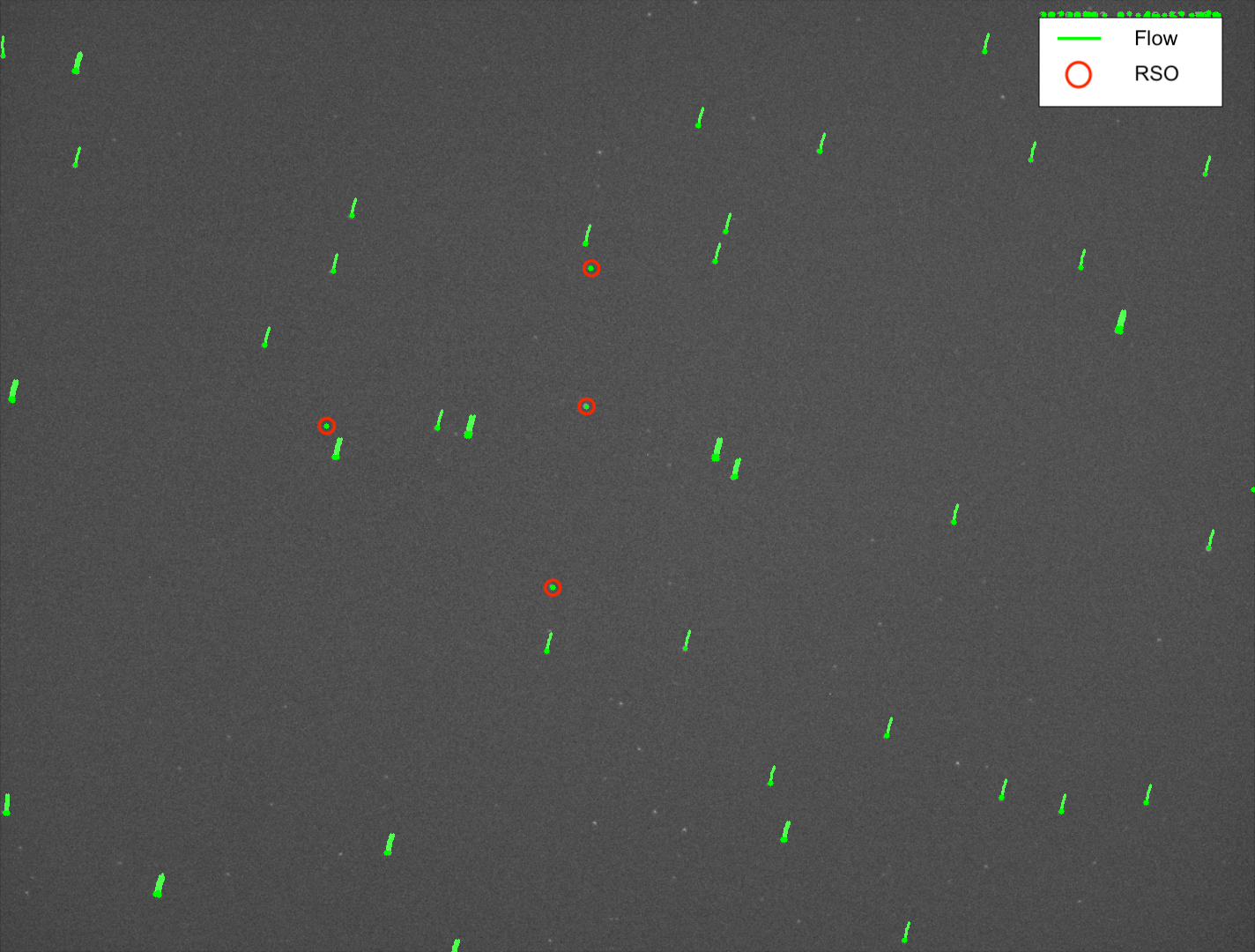}
        \caption{Optical flow RSO detection\newline from ground imagery.}
        \label{fig:groujnd:optical:flow:opencv}
    \end{minipage}
\end{figure}

Figure \ref{fig:groujnd:optical:flow:opencv} shows an initial RSO and flow detection from Figure \ref{fig:raw:ground:image:opencv} using the OpenCV pipeline. This set of images represent a different set of images where the telescope was set to track the satellite motion (GEO satellites) and stars streaked through the image. The main image motion then occurs for the stars. Results are comparable to the MATLAB implementation in that detection was reasonably accurate across the image set.

OpenCV was able to track successfully 4 out of 5 RSOs, with one in particular not being tracked due to the low surface brightness of the satellite. In order to prevent this from happening, image pre-processing similar to the methods employed in the MATLAB pipeline can be implemented. 

\begin{figure}[htb!]
    \centering
    \begin{minipage}[t]{0.24\linewidth}
        \centering
        \includegraphics[width=\linewidth]{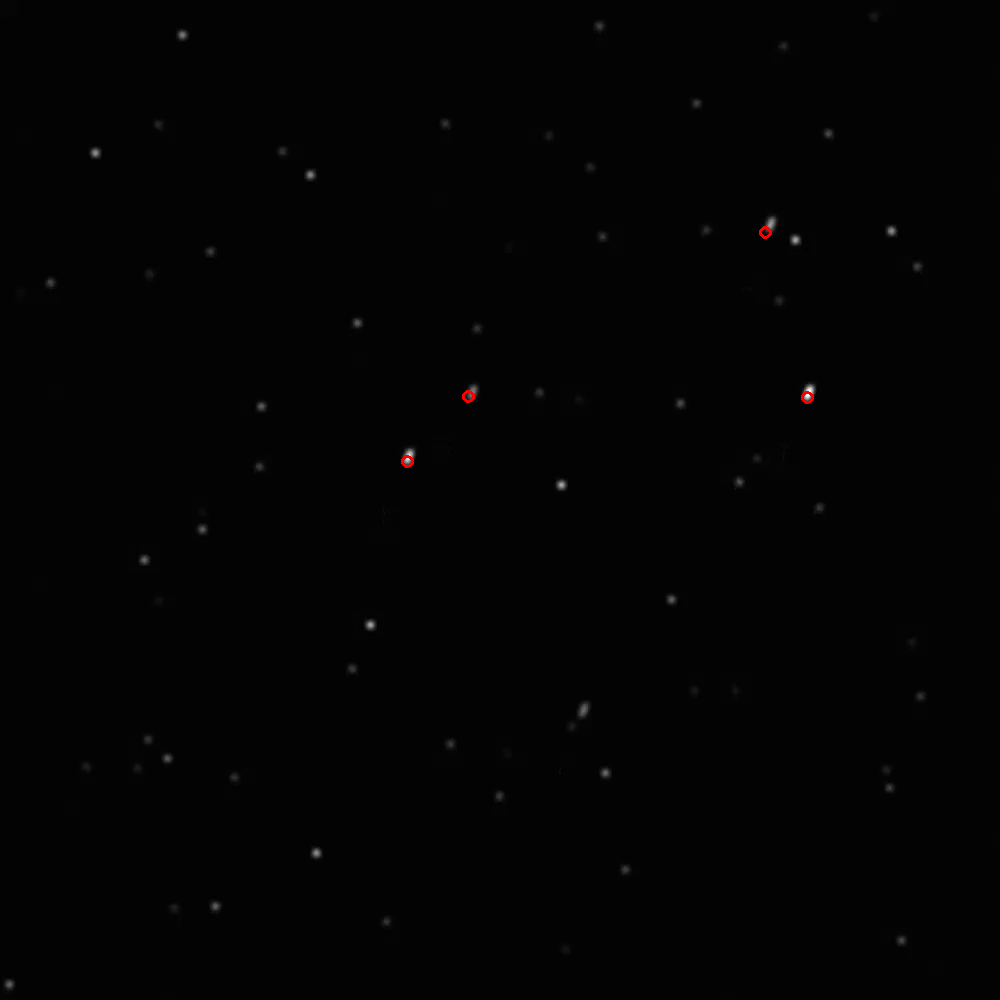}
    \end{minipage}
    \begin{minipage}[t]{0.24\textwidth}
        \centering
        \includegraphics[width=\linewidth]{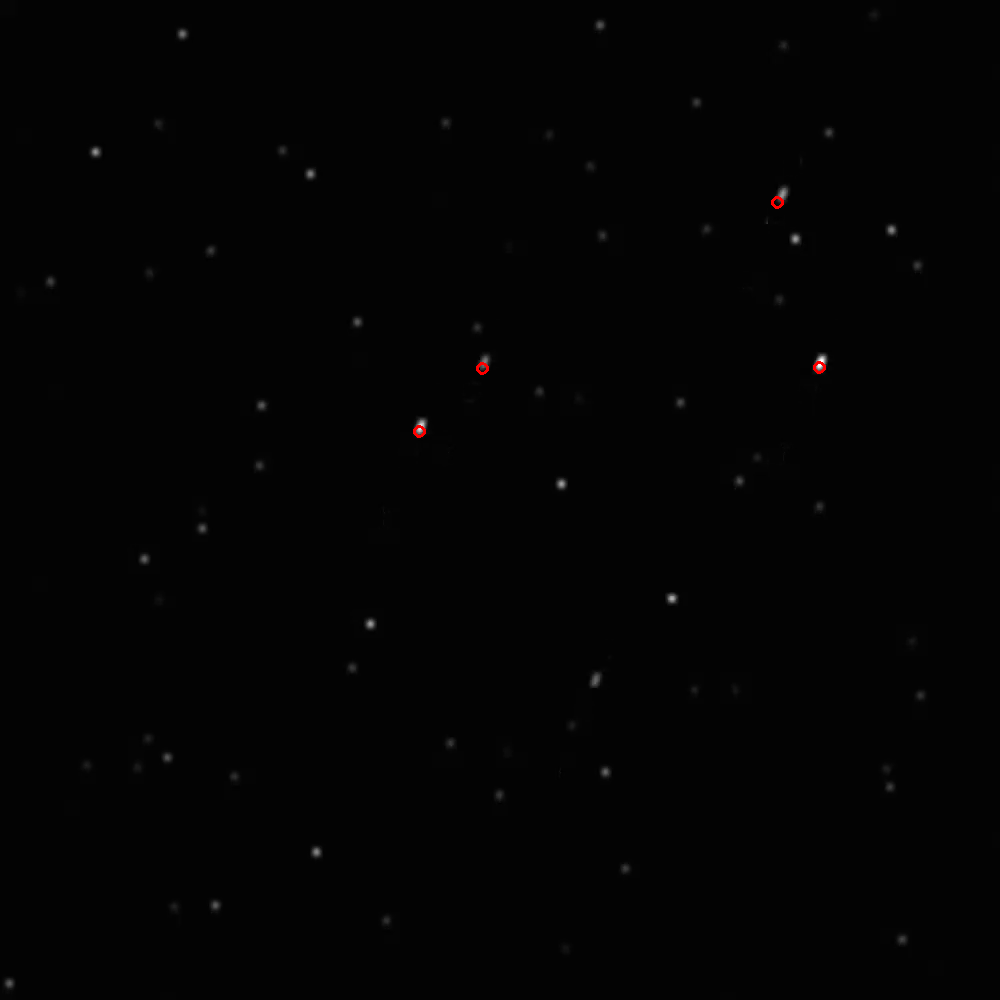}
    \end{minipage}
    \begin{minipage}[t]{0.24\textwidth}
        \centering
        \includegraphics[width=\linewidth]{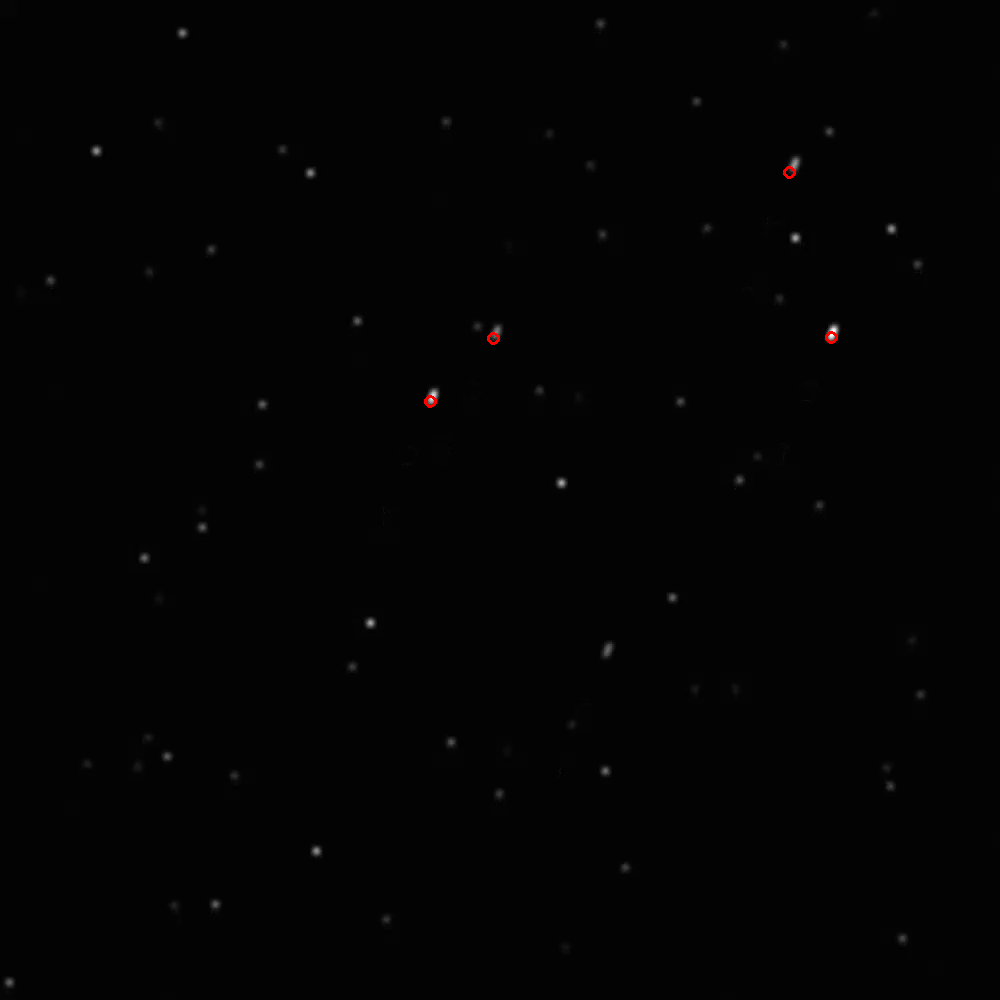}
    \end{minipage}
    \begin{minipage}[t]{0.24\textwidth}
        \centering
        \includegraphics[width=\linewidth]{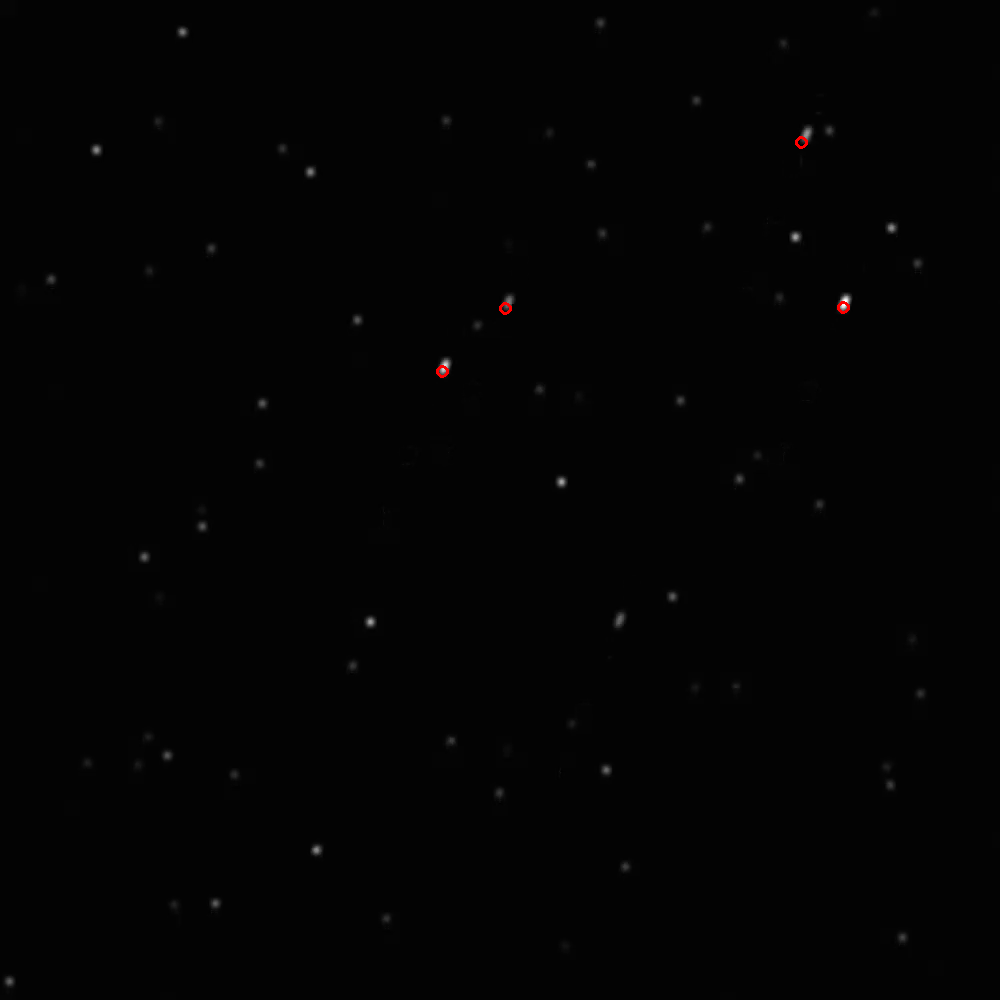}
    \end{minipage}
    \caption{RSOs (marked in red) identified with OpenCV moving from left bottom corner to right upper corner.}
    \label{fig:opencv:rso:detection}
\end{figure}

\section{Conclusion and Future Work}
The autonomous RSO detection algorithm using optical flow was used to successfully track the gross motion of stars and RSO objects through various simulated and real image sets. Overall image-plane velocity, given by star-motion  proved a successful discriminatory threshold for determining candidate RSO objects. Two separate pipelines utilizing commercial (MATLAB) and open source (OpenCV) implementations of detection algorithms demonstrated promising results for RSO detection. 
OpenCV has proved to be an easy and free alternative to commercial applications for this type of workflow, and even possesses some advantages over MATLAB such as giving the user control of more parameters for optical flow. 
Future work includes improving image pre-processing to maximize detection results, standardizing the algorithm between the two pipelines, and running numerous further image sets to benchmark the detection results. Potential different methods to pre-process the images include blur removal and artifacts due to noise from different sources such as sensors, light, atmospheric effects through advanced image filtering.

\bibliographystyle{AAS_publication}
\bibliography{ofsat}

\end{document}